\documentclass{aa}
\usepackage{graphicx}
\usepackage{epsf}

\def\x2{$\chi^{2}$}

\def\rosat{{\it ROSAT }}
\def\asca{{\it ASCA}}
\def\einstein{{\it EINSTEIN }}
\def\sax{{\it BeppoSAX } }
\def\chandra{{\it Chandra} }

\def\lunits{$\rm erg~s^{-1}$}
\def\funits{$\rm erg~cm^{-2}~s^{-1}$}
\def\cunits{$\rm cm^{-2}$}

\begin{document}
   \title{{\it BeppoSAX} Observations of LINER-2 Galaxies}


  \titlerunning{{\it BeppoSAX} Observations of LINER-2 Galaxies}
   \authorrunning{I. Georgantopoulos et al.}

   \author{I. Georgantopoulos\inst{1}
	  \and
          F. Panessa\inst{2,3}
	  \and 
	  A. Akylas\inst{1,4}
	  \and
	  A. Zezas\inst{5}
	  \and
	  M. Cappi\inst{3}
	  \and
	  A. Comastri\inst{6}
          }

   \offprints{I. Georgantopoulos, \email{ig@astro.noa.gr}}

   \institute{Institute of Astronomy \& Astrophysics,
              National Observatory of Athens, 
 	      Palaia Penteli, 15236, Athens \\
         \and
              Dipartimento di Astronomia, Universita di Bologna, 
		Via Ranzani 1, 40127, Bologna\\
	 \and 
		Istituto Technologie e Studio delle Radiazioni 
	Extraterrestri/CNR, via Gobetti 101, I-40129, Bologna \\
        \and 
	 Physics Department, University of Athens, Panepistimiopolis, 
	 Zografos, 15783, Athens \\
	\and 
	Harvard-Smithsonian Center for Astrophysics, 60 Garden St., 
	Cambridge, MA 02138, U.S.A. \\	                        
	\and 
	Osservatorio Astronomico di Bologna, via Ranzani 1, 
	I-40127, Bologna \\
             }

   \date{Received ; accepted }

   \abstract{
 We present {\it BeppoSAX} observations of 6 ``type-2'' LINER and ``transition'' 
 galaxies (NGC3379, NGC3627, NGC4125, NGC4374,  NGC5195 and NGC5879)
 from the Ho et al. (1997) spectroscopic sample of nearby galaxies.
 All objects  are detected in the 2-10 keV band, 
 having luminosities in the range 
 $L_{\rm 2-10 keV}\sim 1\times10^{39}-1\times10^{40}$ \lunits.
 The PDS upper limits above 10 keV 
 place constraints on the presence of a heavily obscured AGN in the case of 
 NGC3379 and NGC4125.  
 No significant variability is detected in any of the objects. 
 The spectra are described in most cases by a simple power-law model 
 with a spectral slope of $\Gamma \sim 1.7-2.5$ 
 while there is evidence neither for a significant absorption above 
 the Galactic nor for an $FeK_\alpha$ emission line. Therefore, 
 based on the spectral properties alone, 
 it is difficult to differentiate between a 
 low-luminosity AGN or a star-forming galaxy scenario. 
  However, imaging observations of NGC3627 and NGC5195 with
 {\it Chandra} ACIS-S  reveal very weak nuclear sources while 
 most of the X-ray flux originates either in off-nuclear 
 point sources or in diffuse emission. 
 The above clearly argue in favour of a star-forming origin 
 for the bulk of the X-ray emission, at least in the above two sources. 
   }

   \maketitle

\section{Introduction}

Low Ionization nuclear emission line regions (LINER) galaxies 
(Heckman 1980) constitute a significant fraction (33 per cent) of 
 nearby galaxies (Ho, Filippenko \& Sargent 1997), and yet 
 the origin of their activity remains under debate. 
 Ho et al. (1997) showed that a small fraction ($\sim$ 20 per cent)
 of LINERs are most probably Active Galactic Nuclei (AGN)
 as they present broad $H_{\alpha}$ emission line wings in their optical spectra.  
 These were classified as LINER-1, while the remaining LINERs
 with no broad $H_{\alpha}$ emission were named LINER-2,   
 in analogy with the existing classification of Seyfert galaxies. 
 
The X-ray emission of LINER-2 provides further clues on the origin of these objects. 
 If LINER-2 are also AGN, in analogy with the 
 obscuration model of Seyfert-2 galaxies (Antonucci \& Miller 1985), 
 their X-ray spectra should show evidence of a
 strong absorption and large equivalent width FeK emission lines. 
 For example the spectrum of the LINER-1.9 NGC1052 (Weaver et al. 1999)
 shows obscuration by a large column density ($N_H\sim 3\times 10^{23}$ \cunits) 
 and an FeK line with an equivalent width of 0.3 keV.  
 In contrast, the {\it BeppoSAX} X-ray spectrum of the LINER-1.9 
 NGC3998 (Pellegrini et al. 2000) 
 shows neither evidence for obscuration nor for an  Fe line. 
 Terashima et al. (2000a) and Roberts, Schurch \& Warwick (2001)
 studied, with {\it ASCA}, about a dozen of type-2 LINER galaxies 
 from the spectroscopic sample of Ho et al. (1997). 
 The  spectra present  slopes with typically $\Gamma\sim 1.8$
 and no evidence for absorption. 
 Terashima et al. (2000a) and Roberts et al. (2001) find 
 no strong evidence for the presence of an FeK line.
 Furthermore, Terashima et al. (2000b) find that 
 if LINER-2 are low luminosity AGN (LLAGN), their X-ray luminosities 
  are insufficient to power their $H_{\alpha}$ luminosities. 
 This suggests an extra source of ionizing radiation (possibly hot stars) 
 or that the AGN, if present, are Compton thick 
 i.e. they are obscured even in the 2-10 keV energy band.   
 In the latter case Terashima et al. (2000b) postulate that 
 we may be observing the scattered component of the 
 nuclear X-ray emission in a similar manner to Compton thick Seyfert-2 galaxies    
 such as NGC1068 (Matt et al 2000). 
 Finally, imaging observations of LINER-2 galaxies using \rosat data  
 (eg Komossa et al. 1999, Roberts et al. 2001) show evidence for 
 extension in soft X-ray energies.   
 Terashima et al. (2000a) report evidence for extension 
 in hard {\it ASCA} X-ray images  of NGC4111 and NGC4569 
 suggesting that a large fraction of the hard X-ray emission 
 does not originate from an AGN. 
 More recently, Ho et al. (2001) present snapshot 
 observations of a sample of 24 nearby galaxies  
 containing many LINER-2. Their preliminary analysis 
 which is confined to the {\it nuclear properties}, 
 shows very low levels of nuclear emission in most cases. 

 Here, we expand significantly the samples 
 of Terashima et al. and Roberts et al. by exploring the 
 X-ray spectral properties of 6 type-2 LINERs
 from the spectroscopic sample of 
 nearby galaxies of Ho et al. (1997).  
 We perform spectroscopic and variability analysis 
 using the {\it BeppoSAX} X-ray mission. 
 The majority of the \sax observations are presented here 
 for the first time. 
 Although \asca observations exist for some of our objects,
  the hard X-ray response 
 of the PDS instrument onboard {\it BeppoSAX} gives the opportunity 
 to study the X-ray spectra of LINER-2 galaxies in the 
 ultra-hard ($>$10 keV) X-ray regime. 
 Finally, our {\it BeppoSAX} analysis is 
 augmented by presenting two public \chandra ACIS-S 
 imaging observations of NGC3627 and NGC5195.

\section{The sample}
We selected LINER-2 objects from the 
Ho et al. (1997) spectroscopic sample of galaxies
 which are contained in the public {\it BeppoSAX} database. 
The Ho et al. (1997) sample contains medium resolution,
 high signal to noise nuclear spectra  
of nearby galaxies ($B_T<12.5 $) in the Northern 
 Hemisphere ($\delta>0$). We also included transition objects 
 i.e. emission line nuclei whose optical
 have [OI] strenghts intermediate between those of 
 HII nuclei and LINERS (see table 5 of Ho et al. 1997 
 for the definition of the optical line ratio of the transition objects).
 Ho et al. (1997) propose that the transition objects can be normal LINERS
 whose spectra are diluted by nearby HII regions (but see also Barth \& Shields 2000). 
 Hereafter, we refer to type-2 LINER and transition objects as 
 LINER-2 galaxies. The sample consists of 6 objects.
 Although the \sax MECS and LECS data of NGC3379 and 
 NGC4125 have been previously reported by Trinchieri et al. (2000),
 we re-do the spectral analysis for the sake of uniformity;
 furthermore the variability as well as PDS spectral analysis are presented here
 for the first time.    
 Two more low luminosity active galaxies (excluding Seyfert galaxies) 
 from the Ho et al. sample 
 were found in the public {\it BeppoSAX} database: 
  NGC3998 and NGC4631; NGC3998 is a LINER-1.9 galaxy
 whose {\it BeppoSAX} data have been previously analysed by 
 Pellegrini et  al. (2000) 
 while NGC4631 is a star-forming galaxy  whose 
 {\it BeppoSAX} observations, to our knowledge, have not been presented before. 
 We analyse these two galaxies as well in order to compare 
 their properties with those of LINER-2.   
 Table 1 lists our sample: in columns (2), (3)
 and (4) we give the object classification, distance in Mpc 
  (as given in Ho et al. 1997) and the Galactic column density 
 (Dickey \& Lockman 1990) respectively.
 Throughout this paper we assume a Hubble constant of 
 $H_o=75 ~\rm km~s^{-1}~Mpc^{-1}$. 

\begin{table}
\caption{The Sample}
\begin{tabular}{cccc}
Name & Classification & Distance  & $N_H^{Gal}$             \\ 
     &                &   Mpc   &  $\times 10^{20}$ \cunits \\ \hline
NGC3379 & L2/T2 & 8.1 & 2.8 \\
NGC3627 & T2/S2 & 6.6 & 2.4 \\
NGC3998 & L1.9 & 21.6 & 1.2 \\
NGC4125 & T2 & 24.2  & 1.8 \\
NGC4374 & L2 & 16.8 & 2.6 \\
NGC4631 & HII & 6.9 & 1.3 \\
NGC5195 & L2 & 9.3 & 1.6 \\
NGC5879 & L2 & 16.8 & 1.5 \\ \hline
\end{tabular}
\end{table}

\subsection{Summary of recent X-ray observations}
\subsubsection{NGC3379}
This galaxy has been observed by Roberts \& Warwick (2000)
 and Halderson et al. (2001) 
 with good resolution (5 arcsec FWHM) using the HRI onboard \rosat 
 in the soft X-ray energy band (0.1-2 keV) . 
 Roberts \& Warwick (2000)  detect a single source 
 with $L_x\approx2 \times 10^{39}$ \lunits
 associated with the galaxy nucleus. 
 NGC3379 has also been observed with {\it BeppoSAX} 
 by Trinchieri et al. (2000) 
 as part of a sample of early-type galaxies
 with low $L_x/L_B$ ratios. 
  The {\it BeppoSAX} spectrum is described by a hard Raymond-Smith (RS) 
 component but with large uncertainties ($kT\sim 7^{+17}_{-3}$ keV)
 or with a single power-law with $\Gamma=1.8_{-0.4}^{+0.3}$. 
 
\subsubsection{NGC3627}
 NGC3627 (M66) is an interacting galaxy in the Leo triplet. 
 It has been previously observed by PSPC onboard \rosat
 (Dahlem et al. 1996) showing a luminosity of $L_x\sim 5\times 10^{39}$
 \lunits in the 0.1-2 keV band.  
 NGC3627 has also been observed by {\it ASCA} (Roberts et al. 2001).
 Its spectrum can be fitted by a two component model: 
  a steep power-law with $\Gamma=2.6^{+0.38}_{-0.37}$ plus an RS 
 with $kT=0.88^{+0.02}_{-0.07}$ keV. 
 NGC3627 has recently been observed by {\it Chandra} ACIS-S   
 as part of an X-ray imaging survey of nearby galaxies (Ho et al. 2001). 
  The above authors do not detect a nuclear source in NGC3627, 
 giving an upper limit for its luminosity of  
 $L_x<4\times 10^{37}$ \lunits in the 2-10 keV band, 
 an order of magnitude below 
 the {\it ASCA} luminosity  ($L_{\rm 2-10 keV}\sim 10^{39}$ \lunits)
 in the same band. 
 
\subsubsection{NGC3998}
High resolution \rosat HRI 
 observations of NGC3998 (Roberts \& Warwick 2000) 
reveal two bright X-ray sources: one is associated 
 with the nucleus and one off-nuclear source with 
luminosities (0.1-2 keV) of $L_x\sim4\times 10^{41} $ \lunits 
and $L_x\sim 4\times 10^{39}$ \lunits  respectively.  
 NGC3998 has been observed in hard X-rays by both
 {\it ASCA} (Ptak et al. 1999) and {\it BeppoSAX} (Pellegrini et al. 2000). 
 The {\it ASCA} spectrum is described well by  a single power-law model. 
 The same model described well the {\it BeppoSAX} data up to 100 keV. 
 No lines were detected, at a statistically 
 significant level, by either {\it BeppoSAX} or {\it ASCA}. 
 The {\it ASCA} X-ray luminosity in the 0.4-10 keV band is $L_x\sim 1\times 
 10^{42}$ \lunits. 

\subsubsection{NGC4125}  
\rosat PSPC observations of this galaxy (Fabbiano \& Schweizer 1995)
 show some evidence for extended X-ray emission. 
This galaxy has again been studied by Trinchieri et al. (2000) 
with {\it BEppoSAX} in order to determine the origin of the X-ray 
 emission in an  early-type galaxy sample.   
 Trinchieri et al. (2000) find a two component fit 
 (RS and blackbody) with temperatures of $kT=$ 0.3 and 4 keV 
 respectively. 

\subsubsection{NGC4374}
 Halderson et al. (2001) have obtained both \rosat HRI and 
 PSPC observations of this galaxy (M84). 
 The HRI observations reveal a large fraction of extended emission.  
The X-ray spectrum of this galaxy  has been investigated by  
Matsumoto et al. (1997) using {\it ASCA}. 
 They find a two component RS fit with temperatures of  
$kT\sim10$  and 0.6 keV respectively. 
  Recently, this galaxy has been observed by Ho et al. 
 (2001) and Finoguenov \& Jones (2001) using \chandra. The galaxy has a nuclear luminosity of 
 $\sim 10^{39}$ \lunits in the 2-10 keV band, more than  
 an order of magnitude lower than the  {\it ASCA} luminosity in the same band.
 Instead, the majority of the X-ray emission in the hard band, appears to come from 
 off-nuclear point sources and diffuse X-ray emission (Finoguenov \& Jones 2001). 

\subsubsection{NGC4631}
\rosat HRI observations of Roberts \& Warwick (2001)
 resolve the soft X-ray emission into  two non-nuclear point 
 sources both with luminosities of $L_{\rm 0.1-2 keV}\sim 10^{39}$  \lunits.  
 Joint {\it ASCA} and \rosat PSPC spectral fits (Dahlem, Weaver 
 \& Heckman 1998) identify  at least three different components:
 a  $\Gamma=1.9$ power-law, a thermal ({\sl MEKAL}) component with $kT\sim$ 0.8 keV 
 and a softer with $kT\sim$ 0.2 keV.    

\subsubsection{NGC5195}
This galaxy interacts with NGC5194 (M51). High resolution 
observations with \rosat HRI reveal extended X-ray emission 
 and an off-nuclear
 X-ray source with $L_{\rm 0.1-2 keV}\sim 6\times 10^{38}$ \lunits 
 (Ehle, Pietsch \& Beck 1995, Roberts \& Warwick 2000, 
 Halderson et al. 2001). 
 NGC5195 is connected to M51 with a bridge of diffuse 
 X-ray emission. 
 The \chandra snapshot observations of Ho et al. (2001) do not detect the nucleus 
 giving an upper limit on the 2-10 keV X-ray luminosity of $\sim10^{38}$
 \lunits. 

\subsubsection{NGC5879}
This galaxy has not been detected by \einstein 
 (Fabbiano, Kim \& Trinchieri 1992) yielding 
 an upper limit of $3\times 10^{40}$ \lunits 
 in the 0.2-4 keV band. No pointed \rosat data 
 exist for this source while  
 the {\it ASCA} image is contaminated by the nearby radio-loud QSO 
1508+5714 (Moran \& Helfand 1997). 

\begin{table*}
\begin{center}
\caption{{\it BeppoSAX} Observation Log}
\begin{tabular}{ccccccc}
{\bf object} & {\bf date } & {\bf Sequence No}  & \multicolumn{3}{c}{\bf Exposures (ksec)}
 & Off-axis (arcmin)     \\ 
     &    &       & MECS &  LECS & PDS  & \\ \hline 
NGC 3379 &  1998-12-14 & 005474 & 98.4 & 35.1 & 50.3 & 0 \\           
NGC 3627 &  1998-12-19 & 005488 & 18.6 & 18.6 & 22.4 & 0 \\                    
NGC 3998 &  1999-6-29  & 006333 & 76.9 & 24.9 & 37.9 & 0 \\                       
NGC 4125 &  1997-4-26  & 002284 & 57.2 & 22.2 & 22.2 & 0 \\
NGC 4374 &  1999-01-22 & 005519 & 67.9 & 66.3 & 51.2 & 15 \\                       
NGC 4631 &  1997-12-18  & 003469& 97.0 & 15.0 & 20.8 & 0  \\                       
NGC 5195 &  2000-1-20  & 007183 & 98.9& 37.2   & 44.9   & 2  \\                  
NGC 5879  &  1998-2-1  & 003751 & 17.9 & 7.0 &  4.2
      & 5  \\             
\hline
\end{tabular}
\end{center}
\end{table*}     

\section{Observations}
The scientific instrumentation on board the Italian-Dutch X-ray  Satellite 
{\it BeppoSAX} includes a Medium Energy Concentrator Spectrometer, MECS, 
which consists of three units, (Boella et al 1997) 
a Low Energy Concentrator Spectrometer, LECS, (Parmar et al 1997) 
a High Pressure 
Gas Scintillation Proportional Counter, HPGSPC, (Manzo et al 1997)
and a Phoswich Detector System, PDS, (Frontera et al 1997), all of which 
 point in the same direction.
The MECS instrument consists of a mirror unit plus a gas scintillation 
proportional counter and has imaging capabilities. It covers the energy range 
between 2-10 keV with a spatial resolution of about 1.4 arcmin at 6 keV and a spectral 
resolution of 8 per cent at 6 keV. 
The three different (or two after May 7 1997, when MECS1 failed) MECS units
are merged  in order to increase the signal-to-noise ratio. This is 
feasible because the three MECS units show very similar performance and  
the difference in the position of the optical axis in the three units is 
smaller than the scale on which the vignetting of the telescopes varies 
significantly ($>$5 arcmin). 
The LECS instrument is similar to the MECS
but operates down to energies of 0.1 keV. 
 The PDS is a direct view detector with 
rocking collimators and extends the {\it BeppoSAX} 
 bandpass  to high energies (13-300 keV).  Its  energy resolution is 15 per cent at 60 keV. 
  Finally, HPGSPC detects photons with energies up  to 120 keV 
  and has an energy resolution of 4 per cent at 60  keV. 

  Table 2 lists the exposures  
  times and Sequence Number for  all observations. 
  Note that all objects apart from  three 
  (NGC4374, NGC5195 and NGC5879), were  the 
  observation targets and therefore were  
  observed on-axis. The off-axis angles  for each object 
  are listed in the last column of table 2.  
  All objects have been detected in the 2-10  keV band by MECS.
  We have not used the PDS data in the case of  NGC4374, NGC5195 
  and NGC5879 as there are bright contaminating  
  sources (the observation targets) within the PDS  field-of-view.
  Only NGC3998 is significantly detected with the PDS  while 
  no object is detected by the HPGSPC.

\section{Short Scale Variability Analysis}
 
 The MECS light curves were examined for evidence
 of short term X-ray variability.   
 We use a binning time of 5 ks. 
 The errors correspond to the 68 per cent confidence level. 
 We have used only objects with adequate photon statistics
 ($>100$ counts).  
 Hence, NGC4374 and NGC5879 were excluded from further variability 
 analysis. We fit the light curve with a constant.  
 The resulting $\chi^2$ imply (see table 3) 
 that there is no statistically significant variability in all cases. 
 As an additional check, we have also performed a Kolmogorov-Smirnov (KS) test. 
 The KS method compares the cumulative distribution 
 of the photon arrival times of the source and the background. 
 Again, no variability has been detected in any of the objects at 
 a confidence level higher than 95 per cent.
    
\begin{figure}
\rotatebox{0}{\epsfxsize=9cm \epsffile{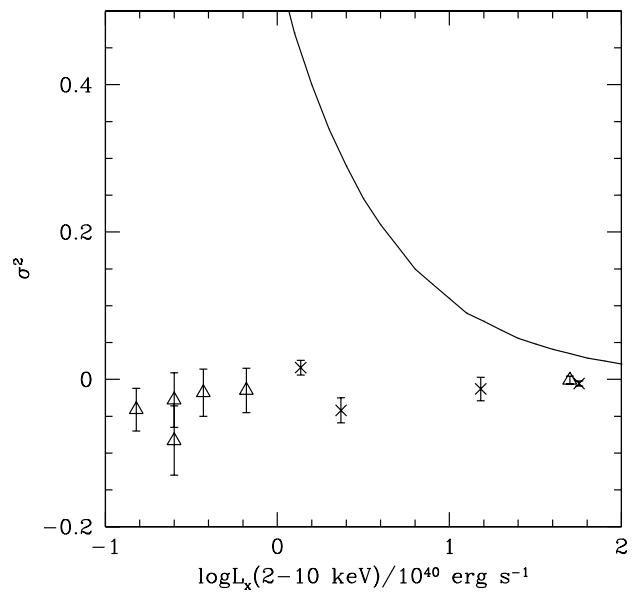}}   
\caption{ The excess variance ($\sigma^2$) versus luminosity (2-10 keV)
 for the 6 LINER objects listed in table 3 (open 
 triangles). LINER galaxies from Ptak et al. (1998) are denoted with 
 crosses. The solid line describes the excess variance for Seyfert  
galaxies derived by Nandra et al. (1997)}
\label{rms} 
\end{figure}

 Finally, we estimate the variability amplitude by means of 
the excess variance $\sigma^{2}_{rms}$ (for a definition of this quantity see 
Nandra et al. 1997).       
 In Fig. \ref{rms} we plot the excess variance as a function of the 
 unobscured 2-10 keV luminosity. We compare the  variance  of the LINER galaxies in our sample 
 to that expected for low luminosity Seyfert galaxies: the solid line represents the 
 extrapolation at low luminosities of the variance-luminosity relation found 
 by Nandra et al. (1997). On Fig. \ref{rms} we also present  4 LINER galaxies 
 from the sample of Ptak et al. (1998). It is evident that the LINER galaxies do not show 
 the strong variability amplitude which is expected for Seyfert galaxies 
 in the same luminosity range.

\begin{table}
\begin{center}
\caption{ Variability Analysis}
\begin{tabular}{cccc}
Name & count rate & $\chi^2$ & $\sigma^{2}_{rms}$   \\
     &  $\times10^{-2}$ &  & $\times10^{-3}$  \\ \hline 
NGC 3379 &  0.3  &38/47 & -41$\pm$30    \\     
NGC 3627 &  0.8  &13/20 & -28$\pm$38     \\     
NGC 3998 &  13.4 &26/33 & -1 $\pm$5        \\     
NGC 4125 &  0.2  &17/17 & -15$\pm$30      \\     
NGC 4631 &  1.0  &11/18 & -18$\pm$33      \\     
NGC 5195 &  0.4  &19/37 & -80$\pm$47      \\    
\end{tabular}
\end{center}
\end{table}

\section{Spectral Analysis} 

 We extract the source spectra 
 using a radius of two arcmin. This area encircles more than  
 85 per cent of the photons at an energy of 3 keV (on-axis).
 The same extraction radii were used in the LECS case. These correspond 
 to a smaller percentage of enclosed energy in the 0.1-2 keV enery range due to the 
 limited spatial resolution of LECS  in the above band.
The spectrum of the  background was estimated from  source free regions of the image.  
We use data between the energy ranges 0.1-4 keV and 2-10 keV  
for the LECS and MECS  detectors respectively 
 where  the response matrices are well calibrated.
The spectral files were rebinned  linearly to give a minimum of 20 counts per channel.   
The spectral fitting was carried out using the {\sl XSPEC} v11 software package.
The MECS and LECS data were fitted simultaneously. A relative normalization factor 
was introduced between the LECS and MECS data. 
 We assumed a  MECS to LECS normalization factor of 
  0.90  to account for cross-calibration 
 uncertainties (Fiore, Guainazzi \& Grandi 1999).

For each observation,  a number of models are applied to the data
 and for each case the $\chi^2$ statistic is estimated.   
We first use a single power-law plus neutral absorption model to fit the spectra. 
Two more complicated models have also been 
 used with the addition of a) a Raymond-Smith (RS),
  plasma model at soft energies; 
 the temperature was constrained to be $kT<1$ keV, while the abundance 
 was fixed to 0.1 in agreement with earlier {\it ASCA} results (Roberts et al. 2001) 
 b) a Gaussian line to account for Fe emission at energies above $\sim$6.4 keV; 
 the line width was fixed to  $\sigma=0.001$ keV i.e. unresolved given the MECS 
 spectral resolution.
   
\begin{figure}
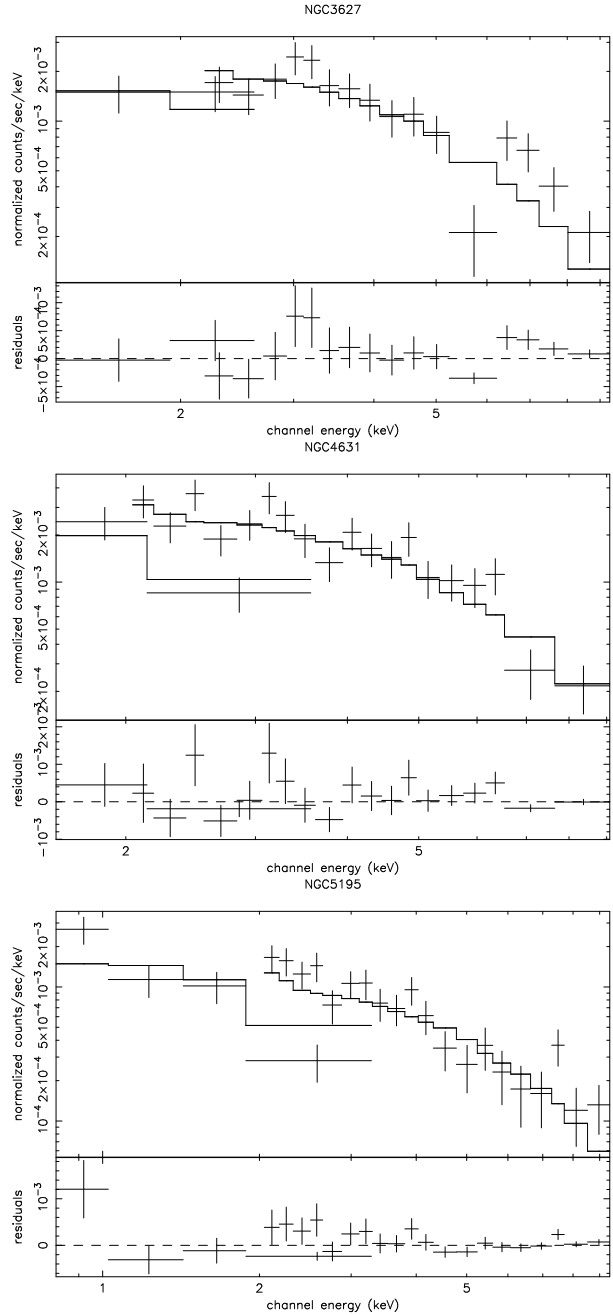

\rotatebox{270}{\includegraphics[height=8.0cm]{3627.ps}}   
\rotatebox{270}{\includegraphics[height=8.0cm]{4631.ps}}
\rotatebox{270}{\includegraphics[height=8.0cm]{5195.ps}}   
\caption{The spectra (data points, 
 together with the best fit models and residuals)
 for the three objects (NGC3627, NGC4631 and NGC5195) whose \sax 
 observations are presented here for the first time  
}
\label{spectra} 
\end{figure}

 The best-fit models are presented in Table 4.
 The best fit spectra together with the residuals for the three 
 objects whose \sax spectra are presented here for the first time 
(NGC3627, NGC4631, NGC5195) are given in Fig. \ref{spectra}.   
 As explained in more detail 
 in the section below, the single power-law model 
 provides a statistically acceptable fit to most objects. 
 In order to assess the significance of new 
 parameters added to the initial model we have adopted the 99 per cent confidence 
 level using the F-test (Bevington \& Robinson 1992). 
 All errors quoted in the best-fit spectral parameters correspond to the 
 90 per cent confidence level for one interesting parameter. 

 Finally, we have used the MECS data and the PDS 
 upper limits in order to derive constraints on the 
 presence of a highly absorbed AGN. 
 In particular, we used only the 15-50 keV PDS upper-limits 
 binned into a single bin, to minimize the background.
 The 15-50 keV 90\% upper limits for NGC3379 and NGC4125 
 are $5\times10^{-2}$ and $6\times10^{-2}$ $\rm cts~s^{-1}$ 
 respectively.  In the case of NGC3627 where \chandra observations show
 that the level of the nuclear X-ray luminosity is much lower 
 than the total \sax emission, (see next section)   
 the PDS cannot provide powerful constraints.  
 We test how our best-fit models (Table 4) for NGC3379 and NGC4125  
 are sensitive to the presence of a heavily 
 absorbed (i.e. $N_H>10^{23}$ \cunits) source. 
 We assume that the normalization of such an intrinsic power-law is 20 times 
 higher than the observed power-law normalization. 
 Indeed, this is a typical value for the ratio of scattered to intrinsic 
 X-ray emission in nearby Seyfert-2 galaxies (eg Turner et al. 1997).
 We further assume the slope of the two power-laws to be 
 identical, but free to vary. Then, by fitting the data,
 we obtain the following 90 per cent confidence lower-limits on the $N_H$ values:
 2 and 1 $\times10^{24}$ \cunits for NGC3379 and NGC4125 
 respectively.

\subsection{Notes on individual Objects}

\subsubsection{NGC3379} 
The single power-law spectrum 
 provides an acceptable fit with $\chi^2=21./16$ degrees of freedom (dof); 
 $\Gamma=1.65^{+0.44}_{-0.30}$ in agreement 
 with previous results by Trinchieri et al. (2000). 
The addition of a RS  component yields  $\Delta\chi^2\sim~ 4$ 
 for two additional parameters
 (normalization and temperature) which does not represent a 
 statistically significant improvement. The addition of 
 a Gaussian line to the single power-law model yields  only $\Delta\chi^2\approx 2.5$ 
 for two additional parameters. 
 The 90\% upper limit for the ew (equivalent width)
 of the line is $\sim$1.5 keV.  

\subsubsection{NGC3627}
 The single power-law fit ($\Gamma=2.35^{+0.55}_{-0.60}$) 
  yields a poor $\chi^2$ (29.8/15 dof) which is 
 rejected at the $\sim$98 per cent confidence level.  
 Still, the addition of a RS component (the temperature 
 was set at 0.7 keV) provides no improvement to the fit ($\Delta\chi\sim 0$). 
 The inclusion of an Fe line ($E=6.78^{+0.32}_{-0.16}$ keV) 
 provides a better fit ($\Delta\chi^2\sim 7$)
 but this is significant at only the $\sim$90 per cent confidence level
 for two additional parameters. 
 The 90 \% upper limit for the Fe line ew is $\sim$2.3 keV. 
 Note that the \asca data (Roberts et al. 2001) favour instead a two component 
 model (power-law plus RS).  
 
\subsubsection{NGC3998}
 This is the observation with the best signal-to-noise ratio.
 The PDS data have been included in the spectral  fits. 
 More detailed spectral fits on the same \sax data were performed 
 by Pellegrini et al. (2000). 
 A single power-law provides a reasonable fit to the data 
 with $\chi^2=300.0/248$. 
 The addition of an RS component ($kT=0.14^{+0.04}_{-0.03}$ keV) is significant at 
 over the 99 per cent conficence level. 
 The power-law slope is $\Gamma=1.88\pm 0.07$   
 while the column density is $\rm N_H=0.3^{+0.1}_{-0.2}\times 10^{22}~cm^{-2}$ 
 i.e. significantly higher than the Galactic column density  of $N_H=0.012\times10^{22}$
 \cunits. 
 
\subsubsection{NGC4125}
 The power-law provides an acceptable fit with $\chi^2=13.5/10$ dof.
 The addition of a RS component ($kT=0.17^{+0.60}_{-0.06}$ keV)
 yields a  better fit with 
 $\Delta\chi^2\sim 4.5$ for two additional parameters; 
 however  this  is 
 statistically significant at only the $\sim90$ per cent confidence level. 
 Both the power-law slope and the column density are poorly constrained:
 $\Gamma=2.52^{+0.58}_{-0.55}$, while the 90 per cent upper limit  
 on $N_H$ is $3\times 10^{22}$ \cunits. Unfortunately, 
 no straightforward comparison can be made with Trinchieri et al. (2000)
 as these authors chose not to fit a power-law model to the data.  
 The addition of an Fe line ($E=6.75_{-0.35}^{+0.25}$ keV)
 is significant at just below the 90 per cent confidence level.  
 The ew of the Fe line is largely unconstrained with its 90\% upper limit 
 being 7.2 keV.

\subsubsection{NGC4374} 
This source was observed at 15 arcmin off-axis. 
It is detected at only the 3.2$\sigma$ level 
having $71\pm22$ counts. 
 Assuming a power-law 
 spectrum with a $\Gamma=1.9$ slope, we obtain a   
 2-10 keV flux of $3.4\times10^{-13}$ \funits,
 corresponding to a luminosity of 
 $\approx1.1\times 10^{40}$ \lunits. 

\subsubsection{NGC4631}
The spectrum is described  by a single power-law model 
 ($\chi^2=22.0/18$ dof) with $\Gamma=2.13^{+0.52}_{-0.44}$
 and no significant evidence for an obscuring column density:
  the 90 per cent uncertainty on $N_H$ is between 0 and 
 $\sim 3\times 10^{22}$ 
 \cunits.  Both the spectral slope observed 
 as well as the absence of a large obscuring column 
 are typical of the X-ray spectra of both 
 type-1 AGN and nearby star-forming galaxies
 (eg Ptak et al. 1997, Zezas, Georgantopoulos \& Ward 1998). 
 Star-forming galaxies present a 
 RS component at soft energies (with typically $kT\sim 0.7 $ keV) due to
  extended hot gas emission.   
  In the case of NGC4631, the addition of a soft thermal component 
  ($kT\approx 0.22^{+0.12}_{-0.10}$ keV)
  does not improve the fit ($\Delta \chi^2\approx 2$),
 in contrast to the {\it ROSAT/ASCA} spectral fits 
 of Dahlem et al. (1998). 
 The addition of a Gaussian line to the power-law spectrum
 is not statistically significant  
 giving $\Delta\chi^2\sim 3.6$; the 90 \% upper limit on the ew 
 is $\sim$1 keV. 

\subsubsection{NGC5195}
  A single power-law model ($\Gamma=1.94^{+0.24}_{-0.21}$) provides a 
 reasonable fit to the data ($\chi^2=34.6/21$ dof).  
 The addition of a soft RS component with a temperature
 fixed at 0.7 keV yields only $\Delta\chi^2=1.6$  
 for two additional parameters. 
 The addition of a line instead gives $\Delta\chi^2=4.3$.
 This feature is marginally significant at only 
 the $\sim$95 per cent confidence level. 
 The line energy is rather high 
 ($E=7.61^{+0.27}_{-0.28}$ keV) while its ew is $\sim$1.3 keV. 
 Interestingly, Pellegrini et al. (2000)
 have also detected such a feature in the case 
 of NGC3998 at an energy  of $7.4^{+0.3}_{-0.2}$ keV  
 at a low  significance level 
 ($<2\sigma$). 

\subsubsection{NGC5879}
 This object is marginally detected at the $\sim3 \sigma$ level.
  We  used a radius of 1 arcmin detection circle 
 to avoid contamination from a nearby ($\sim$ 4 arcmin) quasar.  
 This translates to a flux of $8\times 10^{-14}$ \funits
 or to a luminosity of $3\times10^{39}$ \lunits
 in the 2-10 keV band (assuming a spectrum with $\Gamma=1.9$).   
    
\begin{table*}
\begin{center}
\caption{Best fit spectral models}
\begin{tabular}{cccccc}
Name  &  $N_H$ ($\times10^{22}$ \cunits)  &  $\Gamma$ & $\rm kT$ (keV) & $\chi^2/dof$ & 
$L_{\rm 2-10 keV} (\times10^{40}$ \lunits)  \\  \hline
NGC3379 & $0.05^{+1.05}_{-0.05}$ & $1.65^{+0.44}_{-0.30}$ & - & 21.0/16 & 0.15 \\
NGC3627 & $1.40^{+1.3}_{-1.2}$ & $2.35^{+0.55}_{-0.60}$ & - & 29.8/15 & 0.25 \\
NGC3998 & $0.28^{+0.16}_{-0.11}$ &  $1.88^{+0.07}_{-0.06}$ &  $0.14^{+0.04}_{-0.03}$  & 287.7/248 & 50. \\
NGC4125 & $0^{+0.32}_{-1.6}$ & $2.52^{+0.58}_{-0.55}$ & - & 13.5/10 & 0.66 \\
NGC4631 & $0.9^{+1.7}_{-0.9}$ & $2.13^{+0.52}_{-0.44}$ & - & 22.0/18 & 0.37 \\
NGC5195 & $0.0^{+0.22}$ & $1.94^{+0.24}_{-0.21}$ & - & 34.6/21 & 0.25 \\  \hline
\end{tabular}
\end{center}
\end{table*}     

\begin{figure*}
\vspace{7.0cm}
\caption{The Chandra ACIS-S contours of NGC5195 (left) and NGC3627 (right)
 in the 0.5-8 keV band overlayed on Digital Sky Survey images.
 The diamond and cross represent the position of the 
 UV and radio nucleus respectively. The white vertical bar corresponds 
 to 1 arcmin}  
\label{chandra}
\end{figure*}

\section{{\it Chandra} Imaging analysis}

\subsection{NGC3627}
 In order to study the spatial properties of the X-ray emission of
 these galaxies we used archival data obtained with the
 {\it Chandra} ACIS-S instrument.  Public Chandra data exist only for NGC3627 and 
 NGC5195. These have been observed on 3-11-1999 (exposure 1.1 ksec) 
 and on 23-01-2000 (exposure 1.7 ksec) respectively. 
 Preliminary results from these data have been
 published by Ho et al. (2001), but their study has been focused on the
 luminosity of the nucleus alone. From the raw data we extracted images
 in the 0.5-7.0 keV band which were adaptively  smoothed. Contours from
 these images overlaid on DSS optical images of the galaxies are
 presented in Fig. \ref{chandra}. 
 Source were detected using the wavelet {\sl WAVDETECT}) algorithm 
 of the {\sl CIAO} v.2.0 software package   
 In the case of NGC3627, an X-ray point source  
 close to the radio nucleus (offset 2.6 arcsec)
 is marginally detected. Note that 
 the astrometry error of the \chandra images is usually within 2 arcsec
 and therefore the \chandra nuclear X-ray source may be slightly 
 offset from the radio nucleus.   
  More specifically, 
 the coordinates of the nucleus as derived from radio observations (Filho et al. 2001) 
 are $\alpha=11h20m15.0s$,$\delta=+12d59m30s$ in  J2000
 while those of the X-ray source are $\alpha=11h20m15.1s$,$\delta=+12d59m28s$ (J2000).
 The coordinates of the nucleus derived from UV observations, 
 $\alpha=11h20m15.1s$,$\delta=+12d59m22s$, J2000, 
 (Maoz et al. 1996) are far off from both the radio and the X-ray source  (offset $>6$ arcsec).
 We detect 15 counts from the nucleus translating to an  
  X-ray luminosity of $L_{\rm 2-10 keV}\sim 4\times 10^{37}$ \lunits; 
  we used a radius of 2 arcsec 
  which encompasses over 90 \% of the light at 2 keV from an on-axis point source.  
 For the conversion from counts to luminosities we use a power-law 
 model with $\Gamma=1.9$ absorbed by the Galactic column density.
 Note that Ho et al. (2001) derive an upper limit  
 of $\rm L_{2-10 keV}\approx4\times10^{37}$ \lunits for the luminosity 
 of the {\it radio} nucleus. 
  At least 7 other X-ray point sources are apparently associated with NGC3627. 
 The X-ray emission is dominated by an off-nuclear source 
 (with J2000 coordinates 
 $\alpha=11h20m20.9s$, $\delta=+12d58m45s$) with a luminosity of  
 $L_{\rm 2-10 keV} \sim 10^{39}$ \lunits .
 The  8 point sources (including the nucleus) account for about 60 per cent 
 of the  X-ray emission  within a radius of 2 arcmin in the 0.5-7 keV band. 
 
\subsection{NGC5195}
 In the case of NGC5195 we observe extended X-ray emission 
 peaking $\sim$5 arcsec away from the optical (UV) center of the galaxy.
 In particular, the X-ray emission peaks at $\alpha=13h29m59.5s$, $\delta=+47d15m57s$
 while the UV nuclear coordinates are $\alpha=13h29m59.2s$, $\delta=+47d15m59s$ 
 (Maoz et al. 1996). The coordinates of the nucleus in the radio 
 are $\alpha=13h29m59.5s$, $\delta=+47d15m57s$ according to the NASA Extragalactic 
 Database, (based on VLA observations by Ho \& Ulvestad 2001), coincident with the X-ray peak
 (offset 0.1 arcsec).      
 Given the limited photon statistics it is impossible to 
 discriminate whether this peak corresponds to an additional point source. 
 The nuclear X-ray source has 10 counts (0.5-7 keV) in 2 arcsec region 
 (assuming it is pointlike),   
 translating to a luminosity of $L_{\rm 2-10 keV}\sim 2\times10^{38}$ \lunits.
 Thus it contributes only a small fraction 
 of the total galaxy X-ray emission. The brightest source has 
 a luminosity of $L_{\rm 2-10 keV} \sim 3\times10^{38}$ \lunits 
 while strong diffuse X-ray emission 
 can be clearly seen up to 40 arcsec (1 kpc) with $L_{\rm 2-10 keV}\sim 10^{39}$ \lunits.
 Note that the derived nuclear luminosity in the case of NGC5195 lies above 
 the upper limit of Ho et al. (2001). 
 This discrepancy is probably explained by the different 
 nuclear positions used by us and Ho et al. (2001).  
 Indeed, in the case of NGC5195, Ho et al. (2001) 
 use the optical nuclear coordinates from the 
 POSS plate while we are using the coordinates 
 of the central X-ray source (which we assume to 
 be coincident with the radio nucleus within 
 the errors of the Chandra astrometry).

\section{Discussion \& Conclusions}

 The spectra of all our LINER-2 objects are well described 
 by a single power-law with $\Gamma \sim1.7-2.5$.
 This spectral slope is typical of both LLAGN (Ptak et al. 1999) 
 as well as star-forming galaxies (eg Dahlem et al. 1998).
 Interestingly, our two comparison objects, NGC3998 
 and NGC4631, a bona-fide  AGN and star-forming galaxy 
 respectively according to their high quality optical 
 spectra, again present similar X-ray spectra.
 It is evident that it is quite difficult to
 differentiate between the low-luminosity AGN and the 
 star-forming scenario on the basis of the X-ray 
 continuum alone. The FeK emission line could offer 
 instead a powerful diagnostic. For example the presence 
 of an FeK line at 6.4 keV is frequently used as a definitive 
 proof for the presence of an AGN. Moreover,  
 narrow ionized Fe emission lines due to hot gas around 6.7 keV 
 are more characteristic of  star-forming galaxies 
 (eg M82, Ptak et al. 1997; NGC253, Persic et al. 1998)
  Unfortunately, the limited photon statistics did 
 not allow the detection of any of the above 
 spectral features  in our objects.    
 If the X-ray emission in LINER-2 galaxies emanates mainly 
 from star-forming processes we would expect a strong 
 soft component with a temperature of $\sim0.7$ keV.
   Such a component arises from supernova driven superwinds
 and appears to be  ubiquitous in star-forming galaxies 
 independent of their luminosity 
 (Ptak et al. 1997, Zezas et al. 1998). 
 We found no strong evidence for the presence of a soft 
 component in our objects (apart for the LINER 1.9 galaxy NGC3998
 where the temperature of the RS component is much softer with $kT\sim 0.2$ keV). 
 We believe that a soft component  due to hot gas 
 may be actually present but 
 unfortunately the limited signal-to-noise 
 of the LECS observations coupled with 
 uncertainties in the relative cross-calibration 
 of MECS and LECS hamper its detection. 
  Indeed, {\it ASCA} 
 observations of NGC3627 (Roberts et al. 2001) and NGC4631
  (Dahlem et al. 1998) detect soft X-ray emission with 
 $kT\sim 0.7$ keV in both objects.

\begin{figure}
\rotatebox{0}{\includegraphics[height=8.0cm,width=8.0cm]{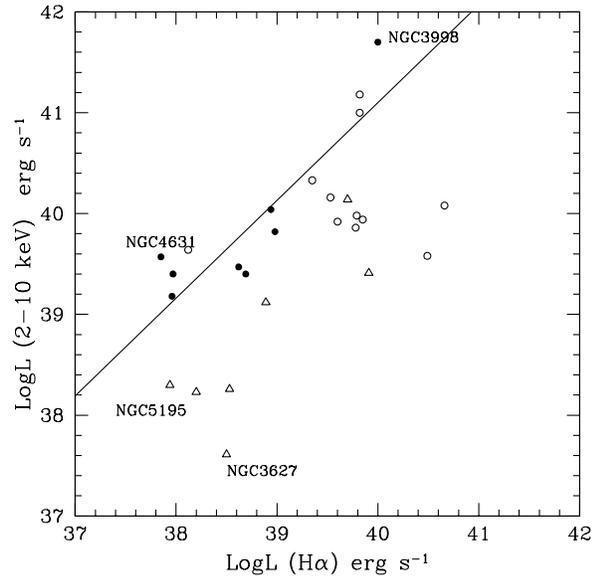}}   
\caption{The X-ray luminosity (2-10 keV) versus the narrow $H_\alpha$ luminosity 
 for the LINER-2 objects in our sample (filled circles), in Terashima 
 et al. (2000b) (open circles). The \chandra 
 nuclear luminosities of NGC3627 and NGC5195 as well as several other 
 galaxies from the sample of Ho et al. (2001) are also shown
  (open triangles). The solid line denotes the $L_x-L_{H_\alpha}$ relation
 holding for type-1 AGN from Terashima et al. (2000b).}      
\label{xha}
\end{figure}

Comparison of the X-ray luminosity and the optical $H_\alpha$ 
 luminosity yields more clues on the  ionizing 
 source in these objects. 
 Indeed, Terashima et al. (2000b) using 
 {\it ASCA} observations of several LINER-2 galaxies
 found that their X-ray emission is insufficient 
 to produce the observed $H_\alpha$ luminosities. 
 According to Terashima et al. (2000b) this means that 
 either an additional ionizing radiation is 
 present (eg hot stars) or that the nucleus 
 is heavily obscured below 10 keV.    
In Fig. \ref{xha} we compare the 
  X-ray luminosities against the $H_\alpha$ 
 luminosities for our sample.  The $H_\alpha$ 
 luminosities were taken from Ho et al. (1997).
 We use the narrow component luminosity correcting for 
 the effects of redenning (see Ho et al. 1997 for details). 
 The objects from Terashima et al. (2000b) 
 are also given on this plot.  
  The solid line gives  the best fit 
 line for type-1 AGN (QSOs, Seyfert-1 and LINER-1) 
 from Terashima et al. (2000b).
 We have also plotted the \chandra {\it nuclear} X-ray luminosities     
 for LINER-2 galaxies from Ho et al. (2001), (detections only),  
 as well as our \chandra nuclear luminosities 
 for NGC3627 and NGC5195. 
 Most of our \sax luminosities follow  the 
 type-1 AGN line. However,  this result appears to be  rather 
 coincidental as all the 
 \chandra nuclear X-ray luminosities lie below the type-1 AGN 
 correlation confirming the claims of Terashima et al. (2000b). 
 It is evident that the large aperture of {\it ASCA} and \sax 
 (a few arcminutes) compared 
 to that used for the optical spectroscopy ($2\times4$ arcsec) 
 alter the true form of the $L_x-L_{H_\alpha}$ relation.  
 The high energy response of the MECS and 
 PDS instruments gives the opportunity 
 to check whether the low X-ray luminosities could be due 
 to high amounts of obscuration. In the case of NGC3379, 
 NGC3627 and NGC4125, we find that 
 the MECS and PDS data are inconsistent with a column density 
 as high as $\sim 10^{24}$ \cunits. If these sources 
 have to be heavily obscured then the obscuring material should have 
 an even larger column and the three sources should
 have to be Compton thick. 
 Alternatively, a more plausible scenario is that hot stars 
 are providing the UV continuum necessary to produce the observed 
 $H_\alpha$ luminosities. Maoz et al. (1995) and Barth et al. (1998) 
 find that about 25 per cent of LINER galaxies display a compact 
 nuclear UV source. In a fraction of these, the {\it HST} UV spectra 
 clearly show absorption line signatures of massive stars 
 indicating a stellar origin for the UV continuum  (Maoz et al. 1998).

 Perhaps, more conclusive evidence on the origin of the 
 X-ray emission in these objects 
 comes from the \chandra imaging analysis.
 The images of NGC3627 and NGC5195 presented here,
 show a very weak nuclear emission 
 with most of the flux arising instead in either 
 off-nuclear sources or in diffuse emission. 
 Similar conclusions are reached on the basis of the Chandra image 
 of  NGC4374  (Finoguenov \& Jones, 2001). 
 These properties are reminiscent of  nearby star-forming galaxies 
 such as M82 and NGC253  (Kaaret et al. 2001, Pietsch et al. 2001,
 Strickland et al. 2000) where the bulk 
 of the X-ray emission originates in off-nuclear sources.    
 Of course it remains to be seen whether the other 
 objects in our sample present similar  imaging 
 properties to NGC3627 and NGC5195. 

In conclusion, the spectral \sax  
 observations  are consistent with both an unobscured LLAGN 
 and a star-forming galaxy scenario. 
 The key test is provided by the 
 \chandra imaging observations which clearly reveal that the 
 a large fraction of the X-ray emission is provided 
 by star-forming processes at least in the 
 case of NGC3627 and NGC5195, where the 
 bulk of the X-ray luminosity has an off-nuclear origin. 
 These observations cannot rule out  
 the presence of a LLAGN in these galaxies.
 However, in this case the  
 nuclear luminosity should be comparable 
 to that of an X-ray binary, unless the nucleus is heavily obscured. 
 Future, longer exposure  
 \chandra and {\it XMM} observations will be able to 
 perform spatially resolved spectroscopy 
 of the nuclear regions.
  The detection of emission lines 
 is expected to provide powerful diagnostics 
 on the nature of the nuclear emission. 

\begin{acknowledgements}
 We are grateful to the referee Dr. Y. Terashima for his 
 numerous comments and suggestions.    
The project was funded by a Greek-Italian scientific bilateral
 agreement under the title ``Observations of 
 active galaxies with the Italian astrophysics mission {\it BeppoSAX}'',
 funded jointly by the Greek General Secretariat for Research and Technology 
 and the Italian Foreign Ministry.  AZ acknowledges support from 
 grant NAS 8-39073. 
 This work has made use of data obtained from the \sax 
 Science Data Center.  
\end{acknowledgements}

\end{document}